\documentclass[referee]{raa}            

\usepackage{graphicx,times}             
\usepackage{natbib}
\usepackage{amssymb,amsmath}
\bibpunct{(}{)}{;}{a}{}{,}

\usepackage{hyperref}
\usepackage{color,colortbl}
\hypersetup{pdftitle = The title of my PDF, pdfauthor = My name,
pdfsubject= The subject, pdfkeywords = keyword1 keyword2 keyword3}
\hypersetup{colorlinks = true, linkcolor = green, citecolor =
blue}
\definecolor{lightRed}{RGB}{230,170,150}

\begin{document}

   \title{New version of the Baade--Becker--Wesselink method based on multiphase effective
   temperature measurements of Cepheids}

   \volnopage{Vol.0 (20xx) No.0, 000--000}      
   \setcounter{page}{1}          

   \author{A. S. Rastorguev\inst{1,2}
   \and Y. A. Lazovik\inst{1,2}
   \and M. V. Zabolotskikh\inst{2}
   \and L. N. Berdnikov\inst{2}
   \and N. A. Gorynya\inst{2,3}
   }

\institute{Lomonosov Moscow State University, Faculty of Physics,
1 Leninskie Gory, bldg.2, Moscow, 119991, Russia;
\\{\it E-mail: alex.rastorguev@gmail.com}\\
\and Lomonosov Moscow State University, Sternberg Astronomical
Institute, 13 Universitetskii prospect, Moscow, 119992, Russia;
\\ \and Instutute of Astronomy RAS, 48 Pyatnitskaya str., Moscow,
119017, Russia\\
\vs \no
   {\small Received~~20xx month day; accepted~~20xx~~month day}}

\abstract{ A new version of the Baade--Becker--Wesselink
(hereafter BBW) method is proposed, based on direct spectroscopic
measurements of effective temperatures of 45 northern Cepheids,
made in different pulsating phases. By comparing the temperature
estimates obtained from the calibration of effective temperature
by normal color with real temperature measurements we were able
not only to determine the color excess with an accuracy of 0.01
mag, but also to derive new color calibration of the effective
temperature immediately for all available measurements with taking
into account the differences in $[Fe/H]$ and $log\,g$ values:
$\log
T_{eff}=3.88-0.20\cdot(B-V)_0+0.026\cdot(B-V)_0^2+0.009\cdot\log\,g
-0.010\cdot(B-V)_0\cdot\log\,g-0.051\cdot[Fe/H]+0.051\cdot(B-V)_0\cdot[Fe/H]$,
which is accurate to about 1.1\%. We also showed the complete
identity of the two main versions of the BBW technique: surface
brightness method proposed by \cite{Barnes+Evans+1976} and maximum
likelyhood method of \cite{Balona+1977}, refined later by
\cite{Rastorguev+Dambis+2010}.
  }

   \authorrunning{A. Rastorguev, Y. Lazovik, M. Zabolotskikh, L. Berdnikov, N. Gorynya }            
   \titlerunning{New version of the Baade--Becker--Wesselink method}  

   \maketitle
\section{Introduction}
\label{sect:intro} Many different approaches have been proposed in
order to solve the problems concerning the universal distance
scale and in particular -- to derive the Period--Luminosity
Relation (PLR) of Cepheids considered as one of most important
``standard candles''. Of course, the trigonometric distances of
Cepheids remain the preferred means of calibration, however, at
present, their trigonometric parallaxes measured during GAIA
mission (\citealt{Gaia}) have significant level of uncertainty and
systematic errors (\citealt{Groenewegen+2018}).

In an absence of reliable and precise trigonometric distances
measured for large sample of Cepheids, BBW method remains one of
the powerful means which allows to independently estimate most
important astrophysical parameters of Cepheid variables --
luminosities and distances -- and to determine the slope and
zero-point of the PLR. In comparison with trigonometric
parallaxes, the distances obtained for Cepheids in open clusters
seems to be more accurate, but limited number of such objects
makes PLR based on Cepheids in open clusters less reliable. The
Baade-Becker-Wesselink (BBW; \citealt{Baade+1926};
\citealt{Becker+1940}; \citealt{Wesselink+1946}) method is thought
now, before GAIA DR3 release, to be more effective and universal
than others approaches. Recent implementations of this method are
the surface-brightness technique (hereafter SB;
\citealt{Barnes+Evans+1976}), maximum-likelihood technique
(hereafter ML; \citealt{Balona+1977}) and its generalization
(hereafter RD; \citealt{Rastorguev+Dambis+2010};
\citealt{Rastorguev+2013}).

A common practice for applying SB technique is to use a linear
calibration of so-called surface brightness parameter with a
normal color. This approach requires a preliminary correction of
the light and color curves for interstellar absorption before
calculating the average radius and luminosity of Cepheid. As for
the original version of the ML technique, it does not require
preliminary correction of photometric data, but it allows you to
determine only the average radius. Note that this method also
implicitly assumes a linear calibrations of the $\log T_{eff}$ and
bolometric correction $BC_\lambda$ with the normal color, as was
noted first by \cite{Rastorguev+Dambis+2010}. The generalization
of ML technique proposed by \cite{Rastorguev+Dambis+2010} uses
nonlinear calibrations of the $\log T_{eff}$ and $BC_\lambda$. An
important advantage of this new approach with well-known nonlinear
approximation for $\log T_{eff}$ and $BC_\lambda$ (see, for
example, \citealt{Flower+1996}; \citealt{Bessell+1998}) is the
possibility to evaluate not only the average radius of the
pulsating star, but also its color excess, flux-averaged absolute
magnitude and the distance. This opens up the independent
possibility to derive the PLR of Cepheids without direct using the
parallaxes.

It is easy to understand that SB technique can be considered as
the simulation of radius changes, while ML technique -- as the
simulation of the light curve. Later in this paper we will show
that both SB and ML approaches are completely equivalent because
they have common physical ground: the Stefan-Boltzmann law and the
relation between measured fluxes, visual and bolometric magnitudes
and distances. This implies that nonlinear approximation of $\log
T_{eff}$ and $BC_\lambda$ is valid for both SB and ML methods.

Now we propose new approach to the problem of Cepheid radii,
extinctions, flux-averaged absolute magnitudes and distances,
based on multiphase effective temperature measurements. The
results of spectroscopic temperature measurements derived by Line
Depth Ratio (LDR) method from high-resolution (R~40000-60000)
echelle spectra were published in a number of papers
(\citealt{Luck+Andrievsky+2004}; \citealt{Kovtyukh+2005};
\citealt{Kovtyukh+2008}; \citealt{Andrievsky+2005};
\citealt{Luck+2008}; \citealt{Luck+2018}). Last paper
\cite{Luck+2018} contains a catalog of 1127 spectroscopic
measurements (effective temperature, composition etc.) made for
435 Cepheids. Most valuable data are given for 52 Cepheids with
five or more temperature measurements.

The main idea comes down to modelling of the effective temperature
curve based on the normal color curve. We used the calibrations of
the effective temperature and the bolometric correction by normal
color (\citealt{Flower+1996}, \citealt{Bessell+1998}). As was
previously shown by \cite{Rastorguev+Dambis+2010}, these two
calibrations provide best fit of Cepheid's observed light curves
simulated by RD method. Due to the sharp dependence of the
effective temperature on the normal color, we were able to
accurately calculate the color excess for all 35 Cepheids. After
that we used 407 multiphase normal colors of these Cepheids with
multiple measurements of the effective temperatures corrected for
the reddening, and derived new calibrations of the effective
temperature by taking into account the differences in $[Fe/H]$ and
$log\,g$.

In this study we use photoelectric and CCD $BV$ photometry of
classical Cepheids from \cite{Berdnikov}, very accurate radial
velocity measurements published by \cite{Gorynya+1992,
Gorynya+1996, Gorynya+1998, Gorynya+2002} and multiphase effective
temperature data taken from \cite{Luck+2018} catalog. We have
selected the sample of 45 Cepheids according to data volume and
quality. We also take into account synchronicity of photometric
and spectroscopic data to prevent any systematic errors in
computed radius and other parameters due to evolutionary period
changes resulting in phase shifts between the light, colour,
temperature and radial velocity curves. As was shown by
\cite{Sachkov+1998} this effect can lead to significant systematic
errors. All calculations were performed by RD method of
\cite{Rastorguev+Dambis+2010} with our new $\log T_{eff}$ --
$(B-V)_0$ calibration and $BC_\lambda$ -- $(B-V)_0$ calibration of
\cite{Flower+1996}.

\section{Physical ground of BBW methods}
\label{sect:Obs}
\subsection{RD version}
As a direct consequence of the Stefan-Boltzmann law and the
relation between absolute and apparent magnitude we can write for
apparent magnitude a generalized Balona's expression (see
\citealt{Rastorguev+Dambis+2010}; \citealt{Rastorguev+2013}) for
details):
\begin{equation}\label{eq1}
  \ m = Y - 5 \cdot \log{\frac{R}{R_\odot}} + \Psi,
\end{equation}
where $Y$ takes a constant value for every Cepheid containing the
apparent distance:
\begin{equation}\label{eq2}
  \ Y = (m - M)_{app} + M_{bol\odot} +  10 \cdot\log{T_{eff\odot}} ,
\end{equation}
and $\Psi$ is a function of the normal color $CI_0=CI-CE$; $CE$ is
the color excess:
\begin{equation}\label{eq3}
  \ \Psi(CI_0) = BC +  10 \cdot \log{T_{eff}} .
\end{equation}

$\Psi$ can be expressed from already available non-linear
calibrations of bolometric correction $BC$ and effective
temperature $T_{eff}(CI_0)$, which also can include terms with
surface gravity ($log\,g$) and metallicity ($[Fe/H]$):
\begin{equation}\label{eq4}
  \ \Psi(CI_0) = a_0 + \sum_{k=1}^{N} a_k CI_0^{k},
\end{equation}

The second component in Eq. (~\ref{eq1}) (used also in SB version)
includes current radius of the Cepheid; its variations can be
obtained by integrating the radial-velocity curve over time taking
into account the projection factor $pf$:
\begin{equation}\label{eq5}
  \ R(\varphi) - R_0 = - pf \cdot \int_{\varphi_0}^{\varphi}(V_r(\varphi) - V_{\gamma} ) \frac{P}{2\pi} d\varphi ,
\end{equation}
where $V_{\gamma}$ is the systematic radial velocity; $R_0$ is the
mean radius value; $P$ is the pulsation period and $\varphi$ is
the current phase of the radial velocity curve. The main
uncertainty of our final results is related to the projection
factor, which determines the ratio of the pulsation velocity and
measured radial velocity. Possible dependence of the projection
factor on the period is still the subject of discussions. In the
present work we used period-dependent $pf$ from
\cite{Nardetto+2007}:

\subsection{SB version}
We start with the expressions for extinction-free illumination
created by star in some photometric band and bolometric
illumination from the star and the Sun:
\begin{equation}\label{eq6}
  \ E = \frac{\pi}{4} \cdot \Phi \cdot \Theta_{LD}^2;
  \ E_{bol} = \frac{\pi}{4} \cdot \Phi_{bol} \cdot \Theta_{LD}^2 = \frac{\sigma}{\pi} \cdot
  T_{eff}^4;
  \ E_{bol \odot} = \frac{\pi}{4} \cdot \Phi_{bol \odot} \cdot \Theta_\odot^2 = \frac{\sigma}{\pi} \cdot T_{eff \odot}^4
\end{equation}
Here $\Phi$ and $\Phi_{bol}$ are surface brightness of the star in
some photometric band and bolometric surface brightness (not
depending on the distance!); $\sigma$ is the Stefan-Boltzmann
constant; $\Theta_{LD}$ is limb-darkened angular diameter of the
star, and $\Theta_\odot$ is Sun's angular diameter. Expressing the
magnitude differences $(m-m_{bol})$ and $(m-m_{\odot})$ in terms
of the logarithm of the illuminations ratios, after some simple
algebraic transformations, we derive an expression
\begin{equation}\label{eq7}
  \ \log \frac{\Theta_{LD}}{\Theta_\odot} = -0.2 \cdot m - 2\cdot F(CI_0) + C,
\end{equation}
where $m$ is extinction-free magnitude,
\begin{equation}\label{eq8}
  \ F(CI_0) = 0.1\cdot BC + \log T_{eff};
  \ C = 0.2\cdot m_{bol \odot} + 2\cdot\log T_{eff \odot},
\end{equation}
Here $F(CI_0)$ is so-called surface brightness parameter
(\citealt{Barnes+Evans+1976}; \citealt{Barnes+2005}) and the
constant $C$ includes Sun parameters. Note that the surface
brightness parameter $F(CI_0) = 0.1 \cdot \Psi(CI_0)$ from Eq.
(~\ref{eq3}).

Taking into account that
\begin{equation}\label{eq9}
  \ \frac{\Theta_{LD}}{\Theta_\odot} = \frac{R}{R_\odot}\cdot\frac{1 AU}{D} =
    \frac{R}{R_\odot}\cdot\frac{1 pc}{D\cdot2.063\cdot10^5},
\end{equation}
and after some algebra we rewrite Eq. (~\ref{eq7}) as
\begin{equation}\label{eq10}
  \ 5\cdot\log \frac{R}{R_\odot} = - m - 10\cdot F(CI_0) + (M_{bol \odot} + 10\log T_{eff \odot} + Mod_0),
\end{equation}
where $Mod_0$ is true distance modulus. In an absence of light
absorption both expressions for RD (Eqs. (~\ref{eq1}-\ref{eq3})
and SB (Eq. (~\ref{eq10})) are completely equivalent up to member
order and designations. It follows from this that SB technique
also requires to use nonlinear calibrations for $log\,T_{eff}$ and
$BC$ by the normal color $CI_0$.

\section{Using multiphase temperature measurements}

\begin{figure}
   \centering
   \includegraphics[width=\textwidth, angle=0]{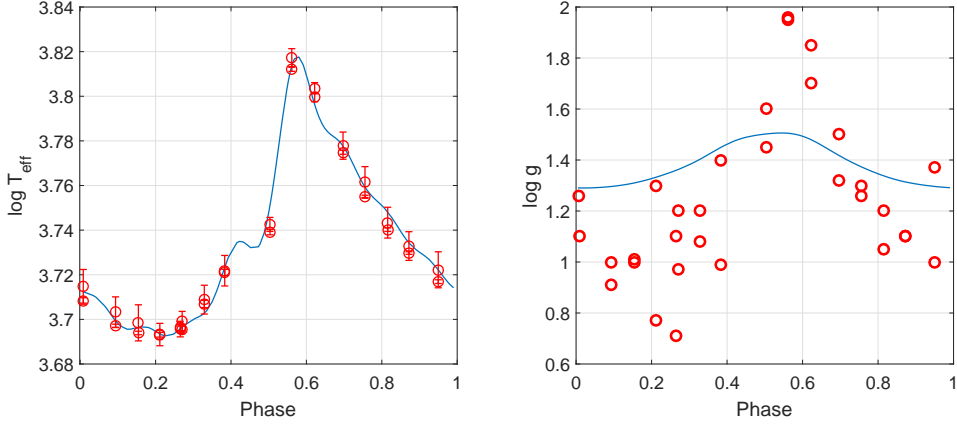}
   \caption{CD Cyg Cepheid. Left: Measured $T_{eff}$ (red circles with error bars) versus
   calculated (blue line). Right: Measured $\log\,g$ values (red circles)
   versus calculated (blue line).}
   \label{Fig1}
\end{figure}

The main point of our new approach is to calculate phase
temperature curve from normal color and radius changes of the
Cepheid using the best calibrations of the effective temperature
and bolometric correction by normal color. We estimate color
excess $E(B-V)$ by requiring the best fit of calculated and
measured effective temperatures. An example of excellent fit for
CD Cyg Cepheid is shown on Fig.~\ref{Fig1} (left panel). As an
example, calculations shown here were performed with calibration
of \cite{Bessell+1998} for the effective temperature with $(B-V)$
color corrected for $E(B-V)\approx0.57\pm 0.005\,mag$. Right panel
shows measured values of $\log\,g$ (red circles) taken from a
number of papers (\citealt{Luck+Andrievsky+2004};
\citealt{Kovtyukh+2005}; \citealt{Kovtyukh+2008};
\citealt{Andrievsky+2005}; \citealt{Luck+2008};
\citealt{Luck+2018}), and the results of our calculations (blue
line) made on the base of direct radius variations (see Eq.
~\ref{eq3}) and Cepheid's mass estimated from Padova evolution
tracks (\citealt{Bressan+2012}). Note unreal large spread of
measured $\log\,g$ values (more than ten times in the surface
gravity!) whereas real radius change is about 20\%. The same is
observed for all stars of our program. The reason for large
variations in $\log\,g$ lies, as it seems to us, in an incorrect
decomposition of the line profiles in the papers mentioned above.

\section{Temperature calibration and Cepheid's luminosities}
\cite{Kovtyukh+2008} (p. 1338) used their temperature measurements
as well as $\log\,g$ values estimated for the sample of 74
non-variable $FGK$ supergiants and multiphase measurements for 164
Cepheids to derive the calibration for normal color $(B-V)_0$ as a
function of $\log T_{eff}, \log\,g, [Fe/H]$ and atmospheric
turbulent velocity $V_t$. It was used by \cite{Kovtyukh+2008} to
estimate the normal colors of supergiants and Cepheids of their
sample with the declared accuracy of $0.050-0.025\,mag$. We guess
that the calibration based on inaccurate $\log\,g$ values cannot
be applicable to calibration of normal colors. For this reason we
derive our own calibration based on normal colors of 32 Cepheids
with best estimations of the $E(B-V)$ by taking into account
calculated variations in $\log\,g$ values. Main steps of our
algorithm are as follows.

\textbf{(1)} By integrating the radial velocity curve according to
Eq. (~\ref{eq5}), we calculate radius variations and then solve
\textbf{RD Eq.(~\ref{eq1})} which allows to find the mean radius.
Now the variations of $\log\,g$ for each star are in hand.
$E(B-V)$ was calculated by requirement of the best fit of measured
$T_{eff}$'s and calculated ones with the calibration by
~\cite{Flower+1996} and \cite{Bessell+1998}(as is shown in
Fig.~\ref{Fig1}).

\textbf{(2)} For each of 32 Cepheids, observed colors $(B-V)$ at
the time of temperature measurements were corrected for color
excess value. Whole set of $(B-V)_0, T_{eff}, \log\,g, [Fe/H]$
values includes 407 points. Based on these data we derived the
calibration for $\log T_{eff}$ as a function of $(B-V)_0, \log\,g,
[Fe/H]$. These calculations were performed for two calibrations:
\cite{Flower+1996} and \cite{Bessell+1998} taken as the first
approximations. Both lead to nearly the same expression $\log
T_{eff}=3.88~(\pm0.01)-0.20~(\pm0.02)\cdot(B-V)_0+0.026~(\pm0.008)\cdot(B-V)_0^2+
0.009~(\pm0.004)\cdot\log\,g-0.010~(\pm0.006)\cdot(B-V)_0\cdot\log\,g
-0.051~(\pm0.017)\cdot[Fe/H]+0.051~(\pm0.022)\cdot(B-V)_0\cdot[Fe/H]$,
which provides relative accuracy about $\sigma_{T}/T\approx1.1\%$.

\begin{figure}
   \centering
   \includegraphics[width=\textwidth, angle=0]{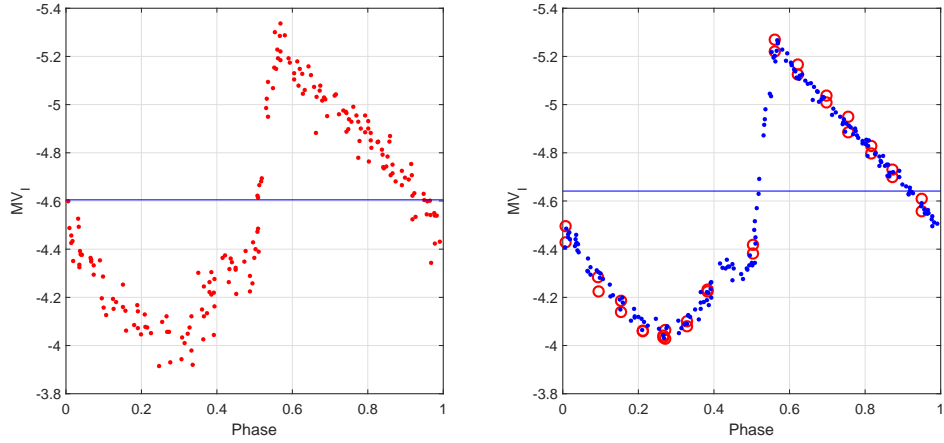}
   \caption{CD Cyg Cepheid. Left: Absolute magnitude curve calculated by corrected $(B-V)_0$ colors.
   Right: The same, but from measured temperatures (red circles). Light-curve shifted by $(V-M_V)_{app}$
   (blue points). Blue horizontal lines show flux-averaged magnitude.}
   \label{Fig2}
\end{figure}

Variations of the absolute magnitude as well as its flux-averaged
value can be calculated by two ways. First, by direct conversion
of corrected colors to temperatures based on our new calbration
(see left panel on Fig.~\ref{Fig2} for CD Cyg). Second, by direct
calculation of luminosity based on measured temperatures (right
panel on Fig.~\ref{Fig2}). In both cases bolometric corrections of
\cite{Flower+1996} and calculated radii variations are used. To
estimate apparent distant modulus we used observed light curve
shifted by $(V-M_V)_{app}\approx 13.56 ~mag$ to fit absolute
luminosity curve calculated by measured temperatures (right panel
on Fig.~\ref{Fig2}). Absolute magnitudes calculated by these two
ways are $<M_V>_I \approx -4.61~mag$ and $<M_V>_I \approx
-4.64~mag$ respectively; which difference can be thought as the
magnitude error.

\begin{acknowledgements}

We thank Russian Foundation for Basic Research (grants nos.
18-02-00890 and 19-02-00611) for partial financial support.
\\

\end{acknowledgements}

\bibliographystyle{raa}
\bibliography{bibtex}

\end{document}